\documentclass[twocolumn]{aastex7}
\usepackage{graphicx}
\usepackage{subcaption}
\usepackage{pgf}
\usepackage{tikz} 
\usepackage{standalone}
\usepackage{adjustbox}
\usepackage{amsmath}
\usepackage{svg}
\usepackage{booktabs}
\usepackage{comment}
\usepackage{gensymb}
\usepackage{booktabs}
\usepackage{multirow}
\usepackage{float}
\usepackage{booktabs}
\usepackage{makecell}
\usepackage{wrapfig}

\begin{document}

\title{Astrometric Parallax Measurements with JWST for Localization of Near-Earth Objects}

\author[0000-0001-6757-3768]{Sriram S. Elango}
\affiliation{Harvard College, Cambridge, MA 02138, USA}
\email[show]{sriramelango@college.harvard.edu}

\author[0000-0003-4330-287X]{Abraham Loeb}
\affiliation{Center for Astrophysics | Harvard \& Smithsonian (CfA), 60 Garden St., Cambridge, MA 02138, USA}
\email[show]{aloeb@cfa.harvard.edu}

\begin{abstract}

We propose the use of the James Webb Space Telescope (JWST) in simultaneous observations with an Earth-based telescope for parallax measurements to tightly constrain the orbital trajectory of hazardous near-Earth objects (NEOs). We demonstrate the significant reduction in localization error with varying epochs of observation at the potential time-of-impact via a Monte Carlo simulated case study of 2024 YR4, an Apollo-type near-Earth asteroid. By leveraging the L2-Earth baseline and the considerable parallax angles formed, we highlight the unexplored potential for improved localization of NEOs through parallax observations with JWST.

\end{abstract}



\section{Introduction}\label{sec:intro}

A massive asteroid impact on Earth carries existential consequences. Asteroids with a diameter $d_a \gtrsim 100  \ \text{km}$ pose the greatest threat to life on Earth (\citealt{SALOTTI2022102933}; \citealt{hawking2018brief}; \citealt{sloan2017resilience}; \citealt{sleep1989annihilation}; \citealt{marinova2011geophysical}; \citealt{napier2015giant}; \citealt{galiazzo2019threat}). Asteroids with sizes comparable to the Chicxulub asteroid impact $d_a \sim 10 \ \text{km}$ can trigger K/T level mass-extinction events with the elimination of $\sim 76\%$ of fossilizable species (\citealt{pope1998meteorite}; \citealt{chapman1994impacts}; \citealt{toon1997environmental}; \citealt{alvarez1980extraterrestrial}). Within the $ 140 \ \text{m} < d_a <  1 \ \text{km}$ range, asteroids can cause substantial localized catastrophes ranging from city-level destruction to significant global effects (climate alteration, earthquakes or tsunamis) (\citealt{chapman1994impacts}; \citealt{rumpf2017asteroid}; \citealt{mathias2017probabilistic}).

More than $90\%$ of asteroids with $d_a > 1 \ \text{km}$ have been detected, thereby lowering the probability of near-term extinction-level events to near-zero (\citealt{NASA}; \citealt{SALOTTI2022102933}; \citealt{morrison1992spaceguard}; \citealt{stokes2003study}). However, less than $50\%$ of asteroids within the $140 \ \text{m} < d_a <  1 \ \text{km}$ threshold have been discovered, failing to meet the $90\%$ threshold outlined by the U.S. Congress (\citealt{NASAJPL_2022};\citealt{NASA_2023}). Even among the known Near Earth Objects (NEOs), many possess considerable uncertainties in orbital trajectories due to observational errors and are challenging to constrain due to their small dimensions and limited albedoes ($\overline{p_{v, \ dark}} \simeq  0.03, \ \overline{p_{v, \ light}} \simeq  0.17$)\footnote[1]{\citealt{nesvorny2024neomod} assume NEO albedo distribution can be approximated using the summation of two Rayleigh distributions with $\overline{p_{v}}$ constants as scale parameters.} (\citealt{Plait_2025}; \citealt{morbidelli2020debiased}; \citealt{nesvorny2024neomod}). Therefore, it is vital that, upon initial detection, hazardous NEOs should be as tightly constrained as possible with minimal positional error to assess impact probability at the time of closest approach to Earth.

The prevailing orbit determination techniques include the Lagrange Planetary Equations (LPEs) and Gauss's Method in which orbital trajectory uncertainties are primarily propagated by errors in distance measurements (\citealt{fernandes2025measuring}; \citealt{milani2010theory}; \citealt{zhai2022role}). Optimizing astrometry observational techniques to minimize error in distance measurements of NEOs achieves the desired objective of orbital trajectory constraints. These distance measurements are acquired by different methods of parallax (traditional geocentric two-observatory parallax, topocentric parallax, rotational reflex velocity [RRV], diurnal parallax), with the latter three methods summarized in detail by \cite{fernandes2025measuring}. 

Traditional geocentric two-observatory parallax utilizes simultaneous observations at two separated Earth-based observatories to measure the angle of the shift of the NEO with respect to the background stars \citep{alvarez2012diurnal}. Topocentric parallax is designed to employ the rotation of the Earth from single (diurnal parallax) or multiple Earth-based observatories at various times, thereby obtaining distinct vantage points (\citealt{fernandes2025measuring}; \citealt{zhai2022role}; \citealt{alvarez2012diurnal}). The Rotational Reflex Velocity (RRV) method requires only one Earth-based observatory and uses the rotation of the Earth and the corresponding angular reflex motion of the asteroids (\citealt{fernandes2025measuring}; \citealt{heinze2015precise}; \citealt{guo2023precise}).

As a consequence of existing parallax methods being Earth-based, distance inferences of NEOs utilize limited baselines $d_b$ (often at most $d_b \approx 2R_\oplus$) and corresponding minor parallax angles, yielding high uncertainties \citep{fernandes2025measuring}. This is no longer the case with the addition of JWST in 2022, which resides at the L2 Lagrange point forming a baseline with Earth of $d_b \approx 235R_\oplus$ (\citealt{giovinazzi2021enhancing}; \citealt{gardner2006james}). Through simultaneous two-observatory parallax, with one observatory being JWST, the other Earth-based, exceptionally significant parallax angles can be formed (depending on the orbital trajectory geometries of the NEO, Earth, and JWST) allowing accurate distance measurements and therefore acquiring tighter constraints on the desired NEO orbit \citep{giovinazzi2021enhancing}. 

Particular high-sensitivity instrumentation within JWST such as NIRCam \citep{beichman2012science} and MIRI \citep{argyriou2023jwst} have already been proven to function at an unprecedented level for asteroid observation and analysis, demonstrating its potential for parallax observations of desired NEOs (\citealt{burdanov2025jwst}; \citealt{rivkin2023near}; \citealt{muller2023asteroids}).

We demonstrate the potential for Earth-JWST parallax in the particular case of 2024 YR4 which was estimated to have a peak Earth impact probability of $3.1\%$ on March 2025 and a $3$ on the Torino scale before that estimate reduced to a near-zero likelihood of impact \citep{Andrews_2025}. The NEO is planned to be observed via JWST during March -- May 2025 and potentially in later years, and is therefore a good representative example of a potential NEO of concern that would benefit from the unique parallax method presented in this study (\citealt{rivkin2025jwst}; \citealt{Josh_2025}).

In this paper, we simulate the proposed simultaneous two-observatory L2-Earth parallax method on the NEO 2024 YR4 to demonstrate the improved localization achieved at the time-of-impact. The parallax observation method and Monte Carlo time-of-impact simulation procedures are documented in Section \ref{sec:methodology}. The localization error and the valid orbits obtained are discussed in Section \ref{sec:results}. Our conclusions and future considerations are summarized in Section \ref{sec:conc}.

\section{Methodology} \label{sec:methodology}

\subsection{Observation Simulation Procedures} \label{sec:observation}

2024 YR4 was detected by the Asteroid Terrestrial-impact Last Alert System (ATLAS) in Chile on December 27, 2024 (\hspace{1sp}\citealt{ATLAS};\citealt{NASA_2025}). As such, we query 2024 YR4, Earth, JWST, and Sun ephemeris vector table data via the NASA JPL Horizons System (referred to as \textit{Horizons}) \citep{NASA_JPL_2025} for the general observation period between December 27, 2024 and June 15, 2025 (approximated date where 2024 YR4 extends beyond JWST visibility). We query \textit{Horizons} for the stated time-frame to acquire vector data (see format in Table \ref{table:jplformatdata}) at a time-step of 1 hour and a defined origin location of the Solar System Barycenter (SSB) (defined as \verb|500@0| in \textit{Horizons}). We utilize only the Julian date (\verb|datetime_jd|) and the respective position vector $\vec{N_t} = [N_x(t),\, N_y(t),\, N_z(t)]$ where $\vec{N}_t$ represents the astronomical body of concern ($\vec{J}_t$ as JWST, $\vec{A}_t$ as 2024 YR4, $\vec{S}_t$ as Sun, $\vec{E}_t$ as Earth) for a time-step $t$.

\begin{table}[htbp]
\caption{\textit{NASA JPL Horizons System Vector Table Data of Earth (399) at $\text{JD} \ 2460671.5$ \citep{NASA_JPL_2025}.} We obtained 4081 rows of the specified format/data detailed for each astronomical object for the selected time frame of December 27, 2024 to June 15, 2025.}
\begin{adjustbox}{center}
  \scalebox{0.7}{
      \begin{tabular}{ccc}
        \toprule
        \textbf{targetname} & \textbf{datetime\_jd} & \textbf{datetime\_str} \\
        Earth (399) & 2460671.5 & A.D. 2024-Dec-27 00:00:00.0000 \\
        \midrule
        \textbf{x (AU)} & \textbf{y (AU)} & \textbf{z (AU)} \\
        -0.09784228839992797 & 0.9742594830258415 & 0.0001274652661564422 \\
        \midrule
        \textbf{vx (AU/d)} & \textbf{vy (AU/d)} & \textbf{vz (AU/d)} \\
        -0.01740724842040341 & -0.001674968390396767 & 4.602493842846188e-07 \\
        \midrule
        \textbf{lighttime (d)} & \textbf{range (AU)} & \textbf{range\_rate (AU/d)} \\
        0.005655157603847823 & 0.9791601860331683 & 7.282898321308521e-05 \\
        \bottomrule
      \end{tabular}
  }
  \end{adjustbox}
  \label{table:jplformatdata}
\end{table}

\begin{figure}
    \centering
    \includegraphics[width=0.9\linewidth]{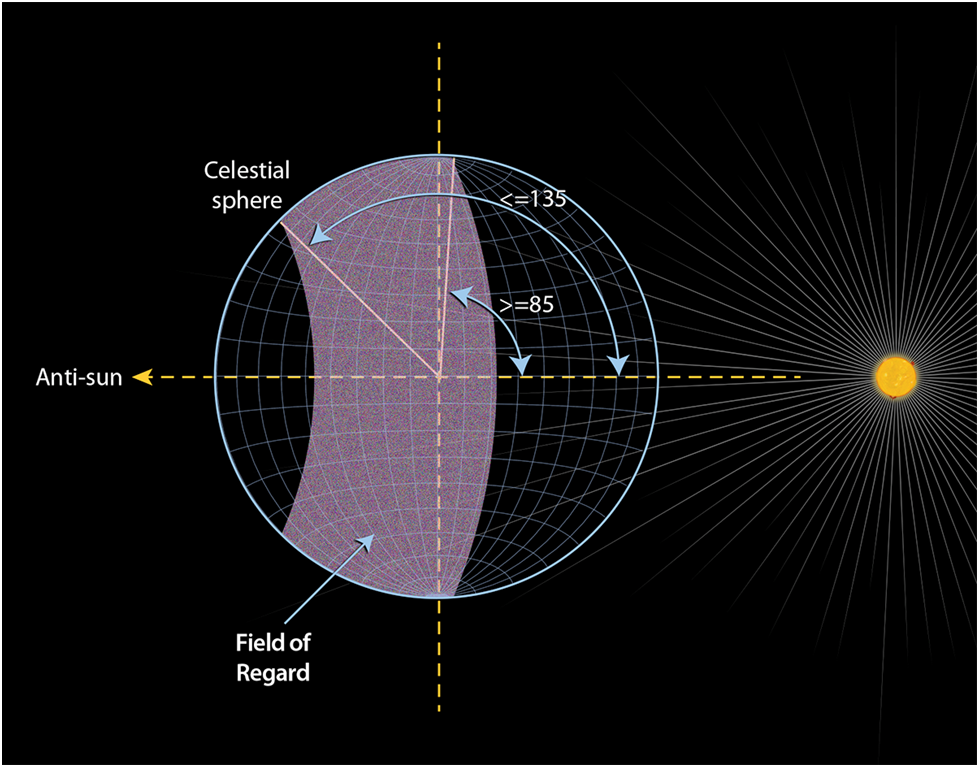}
    \caption{\textit{JWST Celestial Sphere FoR Constraints} (taken from\citealt{JWST_2025b}). JWST can observe astronomical targets safely within the defined range of $85 \degree$ and $135 \degree$ off the Sun-line. Only $\sim 39\%$ of the full sky is observable on any given day \citep{NASAScience_2024}, limiting the opportunity to observe and constrain NEOs. }
    \label{fig:JWSTFOR}
\end{figure}

The selected time frame of December 27, 2024 to June 15, 2025 is not entirely observable by JWST due to its limited Field of Regard (FoR). The JWST FoR is an annulus centered on the position of the Sun, allowing observation of astronomical objects within the range of $85 \degree$ and $135 \degree$ off the Sun-line (see Figure \ref{fig:JWSTFOR} from \citealt{JWST_2025b}) \citep{GoddardSpaceFlightCenter_2006}. 

This constraint must be accounted for within the observation simulation for an accurate representation. We calculate the observable periods of JWST by extracting $\theta_t$ between $\vec{V}_{JA,t}$ and $\vec{V}_{JS,t}$ and identifying if $\theta_t$ satisfies the JWST FoR angle condition $85^\circ < \theta_t < 135^\circ$. We consider observations via JWST exclusively during the time intervals where the JWST FoR angle conditions are fulfilled. 

We further determine the optimal time of each observation for any set of $N$ number of observations by selecting the time with the minimum parallax distance error $\Delta d_t$ defined as,
\begin{equation}\label{eqn:parallaxerror}
\Delta d_t = d_t \times \frac{\delta_{\text{avg}}}{p_t}, \quad \delta_{\text{avg}} = \frac{1}{2}\sqrt{\delta_1^2 + \delta_2^2}
\end{equation}
Here $d_t$ is the true distance to 2024 YR4 (AU), $\delta_1$ and $\delta_2$ are the uncertainties of JWST and an Earth-based telescope respectively (in radians), and $p_t$ is the parallax angle (in radians).

\begin{figure}[h!]
    \centering
    \includegraphics[width=\linewidth]{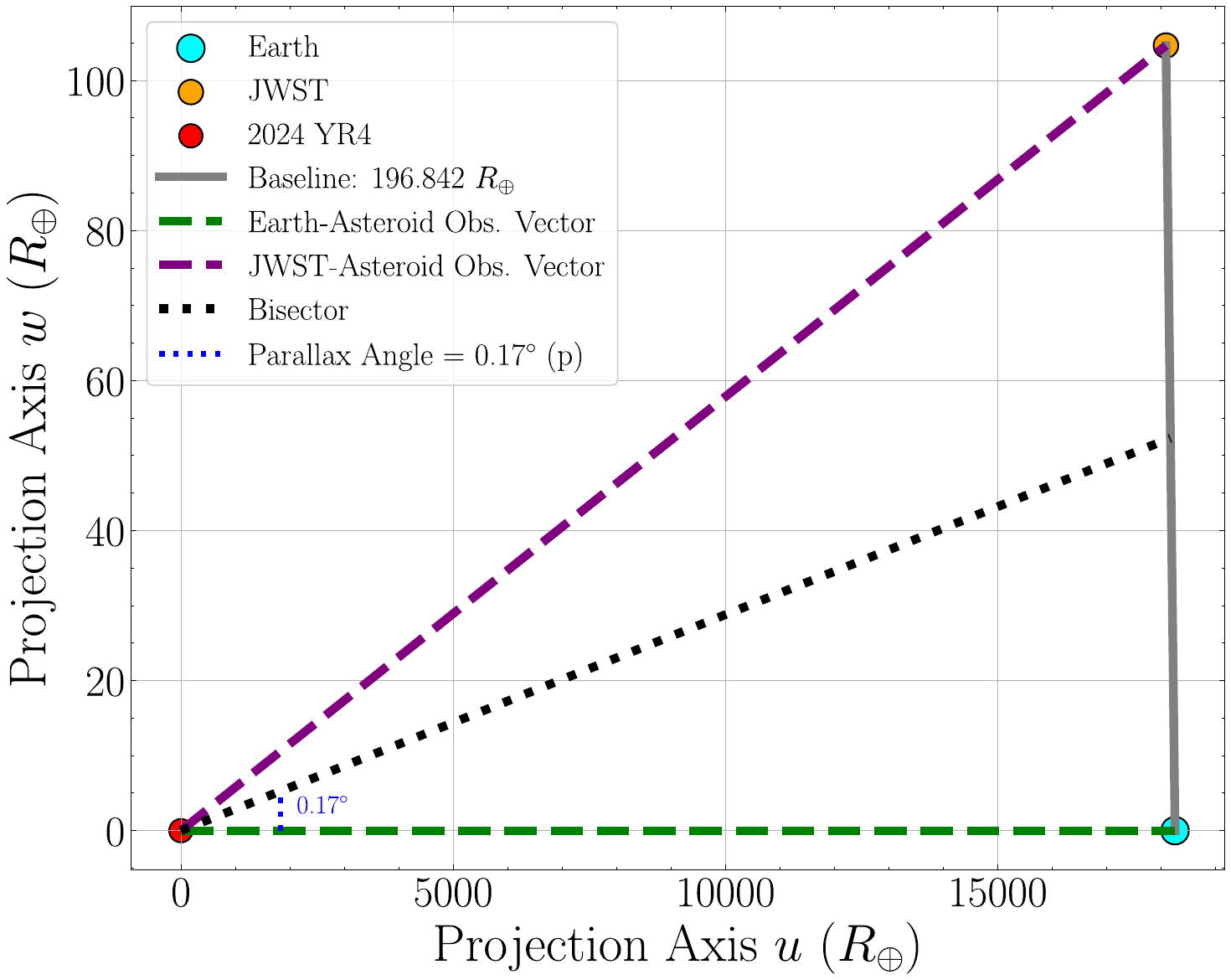}
    \caption{\textit{Parallax Geometry Visualization of Single Observation Epoch}. Plane constructed via the Gram–Schmidt Orthogonalization algorithm with observation vector set $\{\vec{V_{AE}}, \vec{V_{AJ}}\}$. Orthonormal basis formed with $\vec{u}$ as unit vector along $\vec{V_{AE}}$ and $\vec{w}$ as unit vector perpendicular to $\vec{u}$ in the direction of $\vec{V_{AJ}}$. All 3D vectors and points are projected into $(\vec{u}, \vec{w})$ plane (in units of the Earth's radius).}
    \label{fig:parallax_geometry}
\end{figure}

To accomplish this, we first compute the parallax angle $p_t$ for all times $t$ by extracting $\theta_t$ between $\vec{V}_{EA,t}$ and $\vec{V}_{JA,t}$ and defining $p_t = \frac{\theta_t}{2}$. To obtain the true distance $d_t$, we note that the midpoint $\vec{M_t}$ between $\vec{E_t}$ and $\vec{J_t}$ is where $d_t$ is measured from (see Figure \ref{fig:parallax_geometry}). Thus, $\vec{M_t} =\frac{\vec{E_t} + \vec{J_t}}{2}$ and $d_t = \|\vec{A_t} - \vec{M_t}\|$. We define $\delta_1$ as $0.0055^{\prime\prime}$ due to JWST having $5.5 \ \text{mas}$ absolute pointing accuracy with science target acquisition for NIR instruments \citep{JWST_2025a}. We define $\delta_2$ as $0.1^{\prime\prime}$. Converting to the desired units and evaluating equation (\ref{eqn:parallaxerror}) at each time-step $t$ provides a corresponding $\Delta d_t$ for each possible observation time. To obtain the optimal time observation sets with varying $N$ number of observations, we implement an iterative dynamic algorithm with the following conditions: {\it (i)} only utilizing observation epochs within the observable window of JWST and {\it (ii)} enforcing a minimum gap of 7 days between observations. This algorithm then identifies the globally optimal observation time solution sets that minimize $\Delta d_t$ for each observation conducted. Table \ref{tab:optimal_sets} provides the optimal observation times with the respective $\Delta d_t$ for $N$ number of observations between 1 and 8 that satisfy the JWST FoR constraints.\footnote[2]{A maximum of 8 observations are selected as the 8th observation date is the approximated time where 2024 YR4 is too dim for Earth-based (ground) telescopes.}

\begin{table}[ht]
\centering
\caption{\textit{Optimal Observation Sets of 2024 YR4}}
\label{tab:optimal_sets}
\begin{adjustbox}{center}
\scalebox{0.85}{
\begin{tabular}{cccc}
\toprule
\textbf{Set} & \textbf{Observation} & \textbf{Date-Time} & \textbf{Error (\(R_\oplus\))} \\
\midrule
\multirow{1}{*}{1} 
  & 1 & 2025-03-12 00:00:00 & 1.50138923 \\[2mm]
\midrule
\multirow{2}{*}{2} 
  & 1 & 2025-03-08 16:59:59 & 1.51519681 \\
  & 2 & 2025-03-15 16:59:59 & 1.51732833 \\[2mm]
\midrule
\multirow{3}{*}{3} 
  & 1 & 2025-03-07 16:00:00 & 1.52553993 \\
  & 2 & 2025-03-14 16:00:00 & 1.50967560 \\
  & 3 & 2025-03-21 16:00:00 & 1.60461508 \\[2mm]
\midrule
\multirow{4}{*}{4} 
  & 1 & 2025-03-07 16:00:00 & 1.52553993 \\
  & 2 & 2025-03-14 16:00:00 & 1.50967560 \\
  & 3 & 2025-03-21 16:00:00 & 1.60461508 \\
  & 4 & 2025-03-28 16:00:00 & 1.78926146 \\[2mm]
\midrule
\multirow{5}{*}{5} 
  & 1 & 2025-03-07 16:00:00 & 1.52553993 \\
  & 2 & 2025-03-14 16:00:00 & 1.50967560 \\
  & 3 & 2025-03-21 16:00:00 & 1.60461508 \\
  & 4 & 2025-03-28 16:00:00 & 1.78926146 \\
  & 5 & 2025-04-04 16:00:00 & 2.04064868 \\[2mm]
\midrule
\multirow{6}{*}{6} 
  & 1 & 2025-03-07 16:00:00 & 1.52553993 \\
  & 2 & 2025-03-14 16:00:00 & 1.50967560 \\
  & 3 & 2025-03-21 16:00:00 & 1.60461508 \\
  & 4 & 2025-03-28 16:00:00 & 1.78926146 \\
  & 5 & 2025-04-04 16:00:00 & 2.04064868 \\
  & 6 & 2025-04-11 16:00:00 & 2.35704217 \\[2mm]
\midrule
\multirow{7}{*}{7} 
  & 1 & 2025-03-07 16:00:00 & 1.52553993 \\
  & 2 & 2025-03-14 16:00:00 & 1.50967560 \\
  & 3 & 2025-03-21 16:00:00 & 1.60461508 \\
  & 4 & 2025-03-28 16:00:00 & 1.78926146 \\
  & 5 & 2025-04-04 16:00:00 & 2.04064868 \\
  & 6 & 2025-04-11 16:00:00 & 2.35704217 \\
  & 7 & 2025-04-18 16:00:00 & 2.75285531 \\[2mm]
\midrule
\multirow{8}{*}{8} 
  & 1 & 2025-03-07 16:00:00 & 1.52553993 \\
  & 2 & 2025-03-14 16:00:00 & 1.50967560 \\
  & 3 & 2025-03-21 16:00:00 & 1.60461508 \\
  & 4 & 2025-03-28 16:00:00 & 1.78926146 \\
  & 5 & 2025-04-04 16:00:00 & 2.04064868 \\
  & 6 & 2025-04-11 16:00:00 & 2.35704217 \\
  & 7 & 2025-04-18 16:00:00 & 2.75285531 \\
  & 8 & 2025-04-25 16:00:00 & 3.22239554 \\
\bottomrule
\end{tabular}}
\end{adjustbox}
\end{table}

\begin{figure}[H]
    \centering
    \begin{subfigure}[b]{0.9\linewidth}
        \centering
        \includegraphics[width=\linewidth]{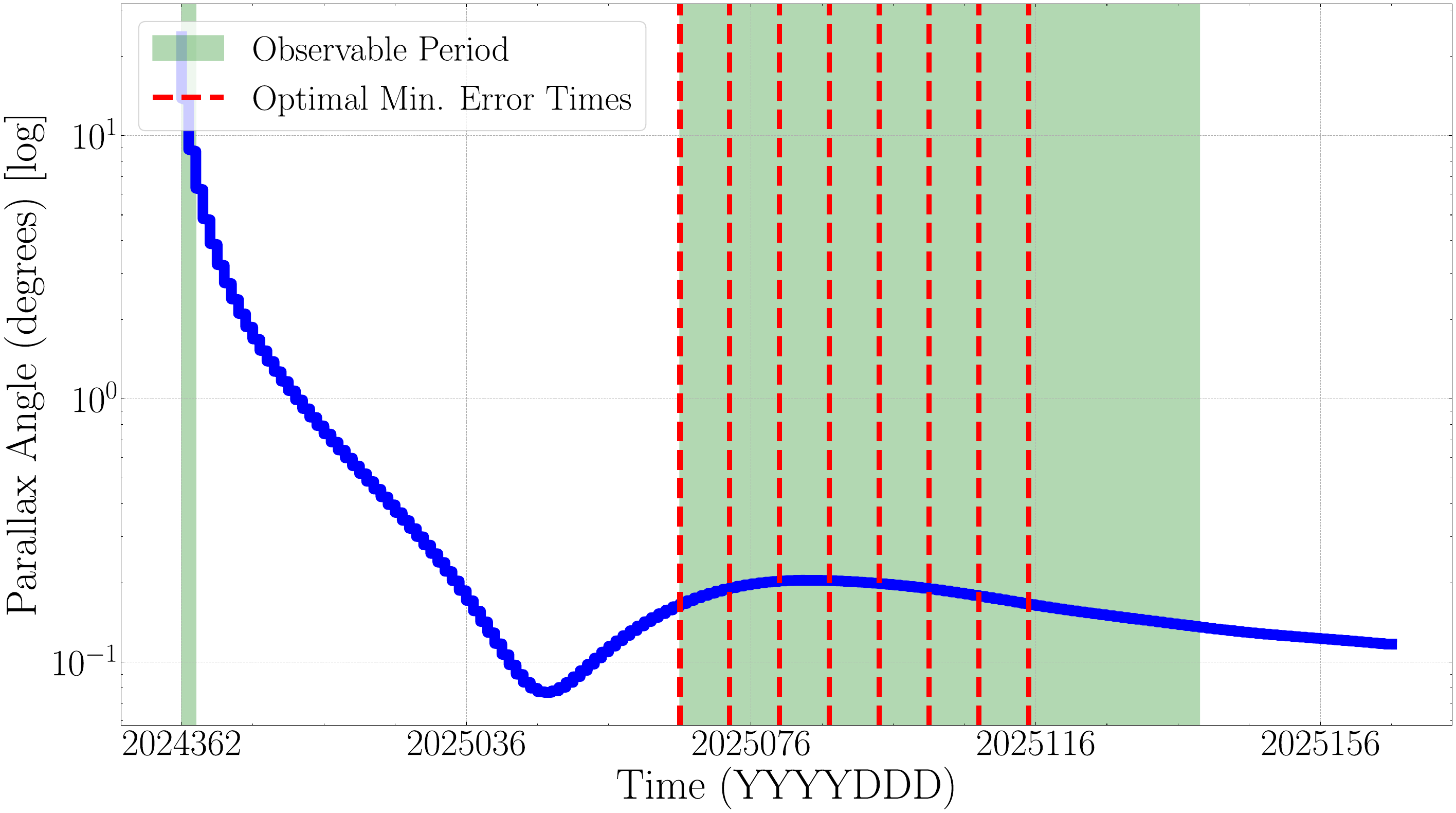}
        \caption{\textit{Parallax Angle $p$ $(\text{degrees})$}}
        \label{fig:angle}
    \end{subfigure}
    \vspace{1em} 
    \begin{subfigure}[b]{0.9\linewidth}
        \centering
        \includegraphics[width=\linewidth]{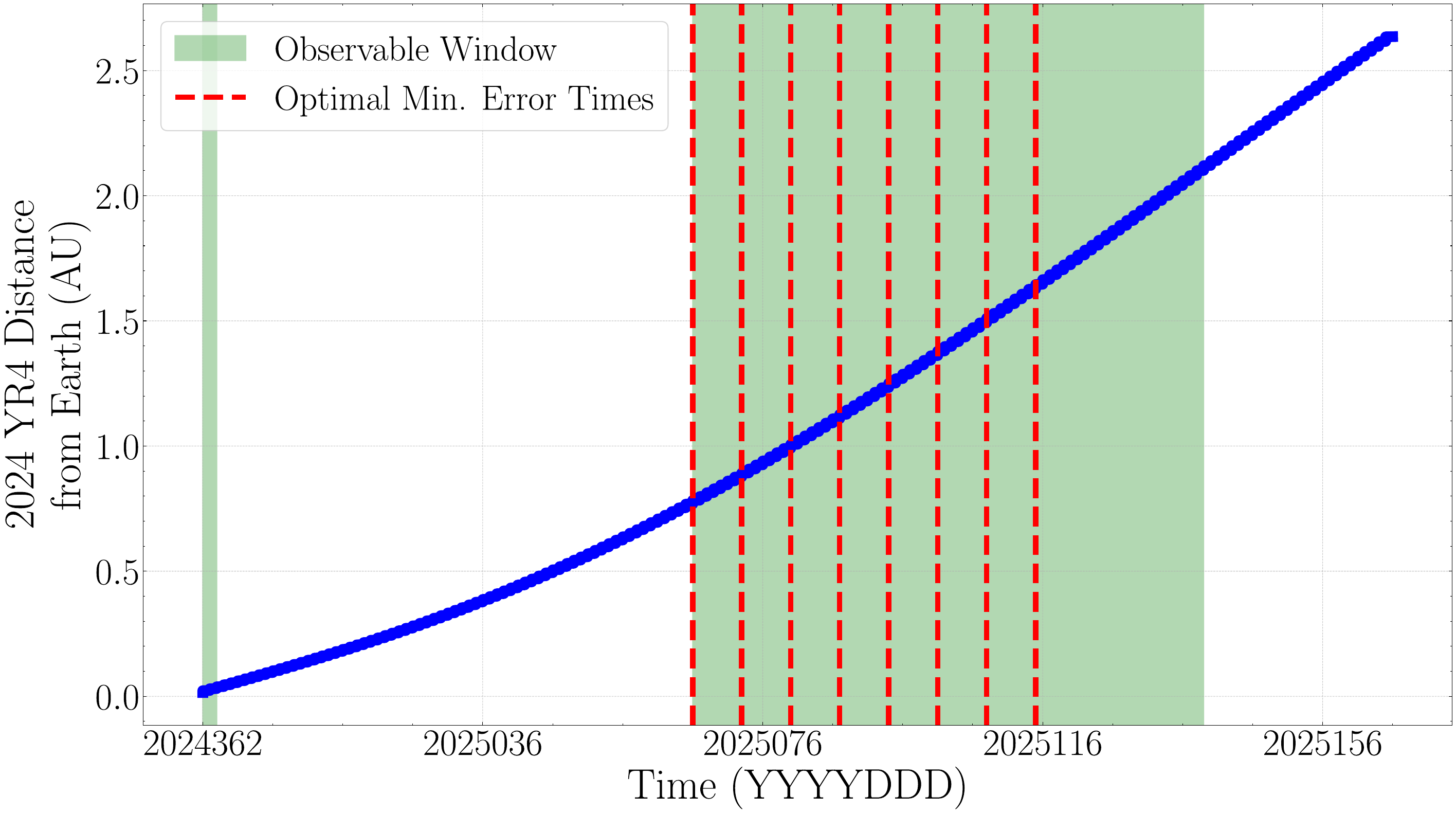}
        \caption{\textit{2024 YR4 Distance from Earth $d_{E,t}$ (AU)}}
        \label{fig:distance}
    \end{subfigure}
    \vspace{1em} 
    \begin{subfigure}[b]{0.9\linewidth}
        \centering
        \includegraphics[width=\linewidth]{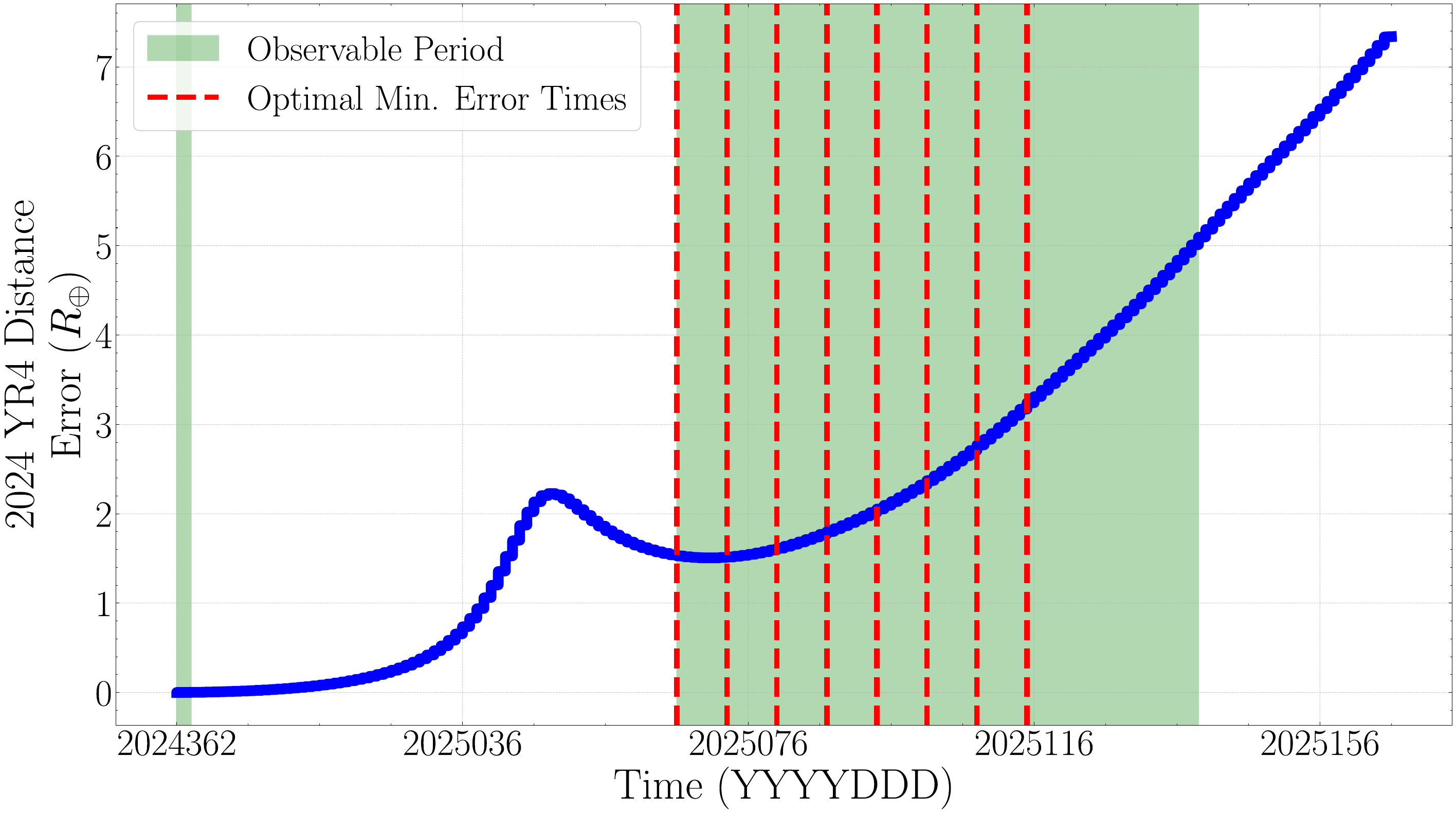}
        \caption{\textit{2024 YR4 Parallax Distance Error $\Delta d_t$ $(R_\oplus)$}}
        \label{fig:distance_error}
    \end{subfigure}
    \caption{\textit{Equation (\ref{eqn:parallaxerror}) Dependent Parameters and $\Delta d_t$ Output for N=8 7-Day Separation Observation Set.} We note that $d_{E,t}$ is not an explicit parameter of equation (\ref{eqn:parallaxerror}), but the explicit parameter $d_t \sim d_{E,t}$. The highlighted green sections are when JWST FoR conditions are satisfied. The optimal minimum error times are indicated in dashed vertical red lines. X-axis is time plotted in YYYYDDD format. Figure \ref{fig:angle} is log-scaled in the Y-axis.}
    \label{fig:2D_analysis}
\end{figure}

We visualize the parallax geometry of a single observation epoch in Figure \ref{fig:parallax_geometry} on the plane formed by the observation vectors $\vec{V_{AJ}}$ and $\vec{V_{AE}}$ with JWST, Earth, and 2024 YR4 being the triangle vertices. We identify a baseline $d_b$ of $\|{\vec{V_{EJ}}}\| = 196.842 R_\oplus$ at this particular time of observation with a parallax angle $p=0.17\degree$. We note, as stated earlier, that the magnitude of the bisector vector formed that divides the triangle at the midpoint $\vec{M_t}$ between $\vec{E_t}$ and $\vec{J_t}$ is $d_t$. The substantial increase in $d_b$ enables a correspondingly significant increase in $p$, which minimizes equation (\ref{eqn:parallaxerror}) and therefore allows tighter constraints on 2024 YR4. The average $\overline p$ possible via the traditional methods of parallax discussed in Section \ref{sec:intro} is vastly smaller due to physical constraints ($d_b \sim 2R_\oplus$) than the example $p$ obtained. We emphasize that $p$ can be magnitudes larger than the scenario demonstrated in Figure \ref{fig:parallax_geometry} depending on the particular geometric relationship in the orbits of Earth, JWST, and the NEO-of-concern (see Figure \ref{fig:angle} for a potential observable period that contains $p \approx 20 \degree$).

\begin{figure}[H]
    \centering
    \begin{subfigure}[b]{0.9\linewidth}
        \centering
        \includegraphics[width=0.9\linewidth]{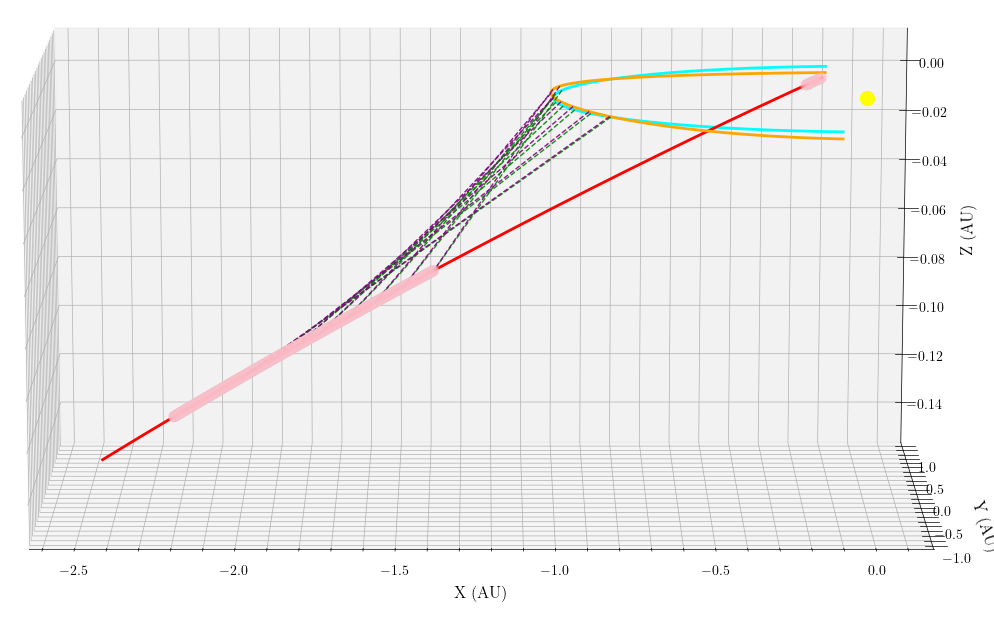}
        \caption{\textit{3D Observation Orbital Geometry from Angle 1}}
        \label{fig:sub1}
    \end{subfigure}
    
    \vspace{1em} 

    \begin{subfigure}[b]{0.9\linewidth}
        \centering
        \includegraphics[width=0.9\linewidth]{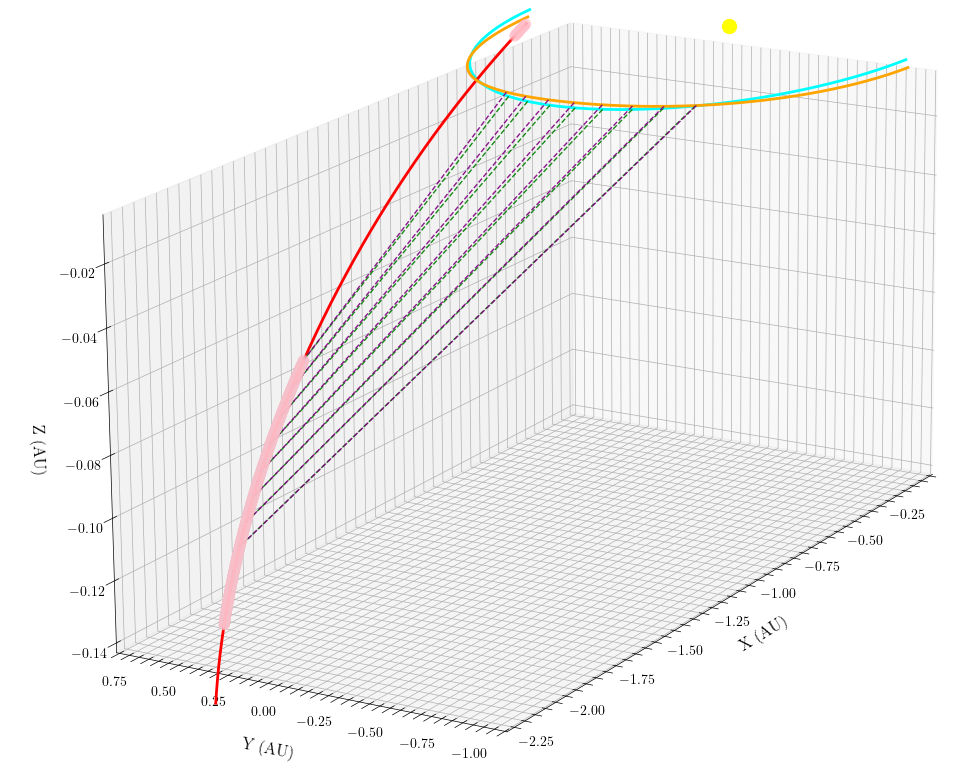}
        \caption{\textit{3D Observation Orbital Geometry from Angle 2}}
        \label{fig:sub2}
    \end{subfigure}
    
    \vspace{1em} 

    \begin{subfigure}[b]{0.9\linewidth}
        \centering
        \includegraphics[width=\linewidth]{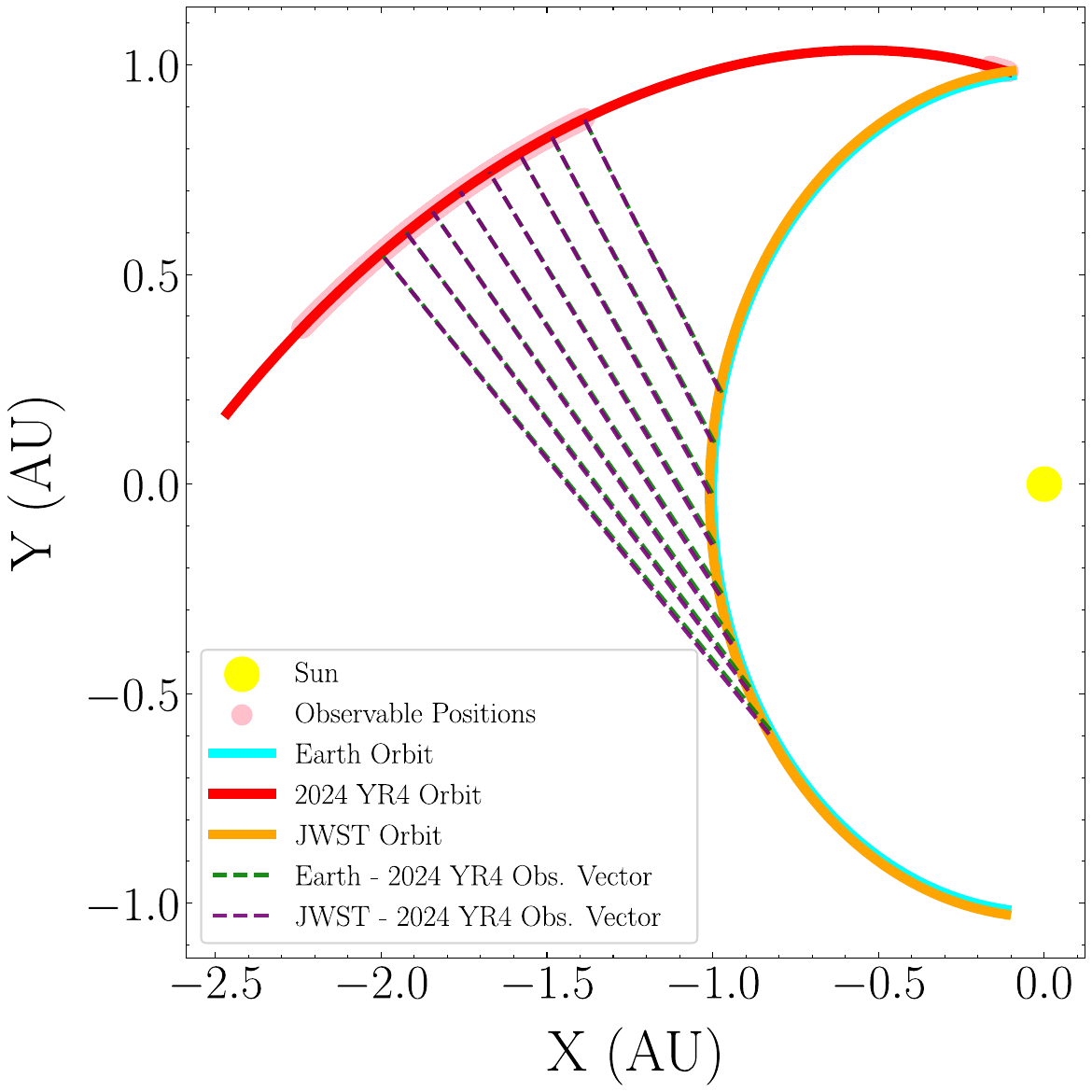}
        \caption{\textit{2D Projected Observation Orbital Geometry on Ecliptic Plane}}
        \label{fig:sub3}
    \end{subfigure}
    
    \caption{\textit{3D and 2D Visualization of $N = 8$ 7-Day Separation Observation Set $\{\vec{V_{EA}}, \vec{V_{JA}}\}$ and $\vec{J_t}$, $\vec{E_t}$, $\vec{A_t}$ Orbital Trajectories.} Plotted for the time frame of December 27, 2024 to June 15, 2025. The global optimal observation time solutions for a set of 8 pairs of observation vectors $\{\vec{V_{EA}}, \vec{V_{JA}}\}$ with 7-day separation is provided in Table \ref{tab:optimal_sets}.}
    \label{fig:orbitviz}
\end{figure}

In Figure \ref{fig:orbitviz} we plot the $\vec{J_t}$, $\vec{E_t}$, $\vec{A_t}$ orbital trajectories, highlighting $\vec{A_t}$ where JWST FoR constraints are satisfied. There are two such regions, however, only the March-May time frame (longer highlighted $\vec{A_t}$ trajectory path) is considered for observations \citep{Josh_2025}. An iterative dynamic program obtains a globally optimal set of 8 observation times with 7-day separation for this period (see Table \ref{tab:optimal_sets}). Parallax observations are conducted at stated optimal times (see Table \ref{tab:optimal_sets}) represented by observation vectors $\vec{V_{EA}}$ and $\vec{V_{JA}}$ in Figure \ref{fig:orbitviz}.

As $\vec{A_t}$ travels to aphelion $(Q)$ away from the JWST-Earth system where $\frac{\partial p}{\partial t}$ is negligible (see Figure \ref{fig:angle}), the minimum $\Delta d_t$ observation times (see Figure \ref{fig:distance_error}) are predominantly dependent on when $\vec{A_t}$ is the closest (see Figure \ref{fig:distance_error}). This behavior often alters when NEOs are in close proximity to the JWST-Earth system where $\Delta d_t$ tends to be dominated by $p_t$ due to rapid changes in $\frac{\partial p}{\partial t}$ at closest approach. We plot in Figure \ref{fig:2D_analysis} the equation (\ref{eqn:parallaxerror}) time-dependent parameters ($p, d_{E,t}$) and output ($\Delta d_t$) for the simulated observation period in Figure \ref{fig:orbitviz}. Upon initial detection of 2024 YR4, an observability window existed with a considerable $p \approx 20 \degree$ and minimal $d_{E,t}$, offering a highly accurate localization. In the second observable window, $\Delta d_t$ had a local minima at $t = 3/12/25$ which could have been exploited for parallax measurements. Optimizing for ideal $p_t$ and $d_t$ values is essential in scheduling observation times for JWST-Earth parallax measurements. 

\subsection{Monte Carlo Time-of-Impact Simulation Procedures} \label{sec:monte}

The projected potential time-of-impact epoch of 2024 YR4 is December 22, 2032 \citep{NASA_2025}. In this section, we implement a Monte Carlo method to compute the localization error (position uncertainty) of 2024 YR4 at the stated epoch for each \textit{N} set of observations (see Table \ref{tab:optimal_sets}) conducted in Section \ref{sec:observation} to calculate the reduction in localization error achieved with each additional parallax measurement.

\begin{table}[ht]
\centering
\caption{\textit{True 2024 YR4 Orbital Elements at December 27, 2024 \citep{NASA_JPL_2025}}}
\label{tab:orbital_elements}
\resizebox{\columnwidth}{!}{%
\begin{tabular}{ccc}
\hline
\textbf{Parameter} & \textbf{Value} \\
\hline
Semimajor Axis ($a$)             & 2.447589641022944 AU \\
Eccentricity ($e$)                & 0.6560404707845989 \\
Inclination ($i$)          & 3.422423121936059$^\circ$ \\
Longitude of Ascending Node ($\Omega$)  & 271.9072914654213$^\circ$ \\
Argument of Perihelion ($\omega$)  & 132.9212426001236$^\circ$ \\
Time of Perihelion Passage ($T_p$)     & 2460636.44097958219 (JD) \\
\hline
\end{tabular}%
}
\end{table}

The true Keplerian orbital values of 2024 YR4 are retrieved via \textit{Horizons} for the initial detection date of December 27, 2024 (see Table \ref{tab:orbital_elements}) and are propagated using a standard analytic Keplerian propagator for the general observation period to obtain the true orbit $\vec{A_{\text{true}}}$. The 2024 YR4 orbital value uncertainties (1-$\sigma$) are further queried (see Table \ref{tab:orbital_uncertainties}) where 10 million perturbed orbital value sets $\{a \pm\sigma, e\pm\sigma, u\pm\sigma, \Omega\pm\sigma, \omega\pm\sigma\}$ are generated via uniform distribution random sampling.

\begin{table}[ht]
\centering
\caption{\textit{One Sigma Uncertainties for Orbital Elements of 2024 YR4 \citep{NASA_JPL_2025}}}
\label{tab:orbital_uncertainties}
\resizebox{\columnwidth}{!}{%
\begin{tabular}{ccc}
\hline
\textbf{Parameter} & \textbf{Uncertainty (1-$\sigma$)} \\
\hline
Semimajor Axis ($a$)             & $ 1.8916\times10^{-5}$ (AU) \\
Eccentricity ($e$)               & $ 2.6997\times10^{-6}$ \\
Inclination ($i$)                & $ 1.71342463\times10^{-7}$ rad \\
Longitude of Ascending Node ($\Omega$) & $ 1.50777249\times10^{-7}$ rad \\
Argument of Perihelion ($\omega$)     & $ 2.29266451\times10^{-7}$ rad \\
\hline
\end{tabular}%
}
\end{table}

\begin{figure*}[t!]
    \centering
    \includegraphics[width=\linewidth]{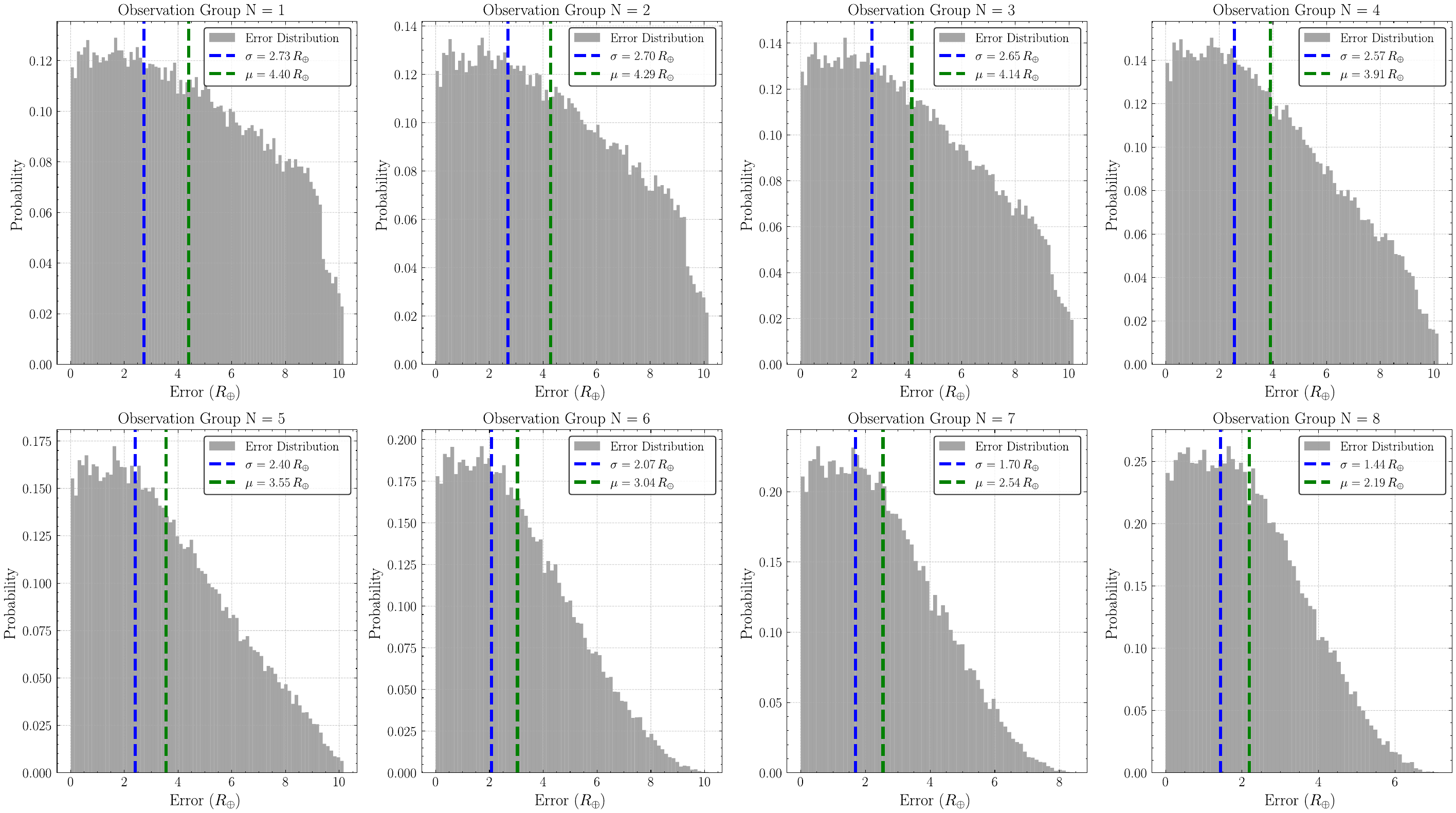}
    \caption{\textit{Histogram of Monte Carlo Valid Perturbed Orbit Deviations $\|\vec{\delta_I}\|$ from True Orbit at Time-of-Impact Epoch for each $N$ Observation Group.} The error on the horizontal axis is normalized by the Earth's radius. Each $N$ observation group plot includes respective standard deviation $(\sigma)$ and mean $(\mu)$ of $\|\vec{\delta_I}\|$ values. The histogram is normalized to form a probability density curve and bin size is determined by the Rice rule.}
    \label{fig:hist}
\end{figure*}

For each \textit{N} set of observations (see Table \ref{tab:optimal_sets}), an effective distance error $\sigma_{\mathrm{eff}}$ is calculated for every observation $i$ conducted within the individual set, where the raw parallax distance error $\Delta d_t = \sigma_{\mathrm{p}}$ is taken as the effective distance error for the first observation $\sigma_{\mathrm{eff},1}$, and every subsequent observation has a calculated effective distance error $\sigma_{\mathrm{eff},i}$ through combining the previous observation effective distance error $\sigma_{\mathrm{eff},i-1}$ with the current observation raw parallax distance error $\sigma_{p,i}$ using the weighted average formula (see equation \ref{eq:efferror}).

\begin{equation}
\begin{aligned} \label{eq:efferror}
\sigma_{\mathrm{eff},1} &= \sigma_{\mathrm{p}} \\
\sigma_{\mathrm{eff},i} &= \frac{1}{\sqrt{\frac{1}{\sigma_{\mathrm{eff},i-1}^2} + \frac{1}{\sigma_{p,i}^2}}}, \quad \text{for } i \geq 2
\end{aligned}
\end{equation}

Each generated perturbed orbital value set is iterated across all \textit{N} sets of observations, where for each observation set, the candidate perturbed orbit $\vec{A_{\text{per}}}$ is propagated to an optimal observation time $i$ (see Table \ref{tab:optimal_sets}) in time-ascending order where a heliocentric vector position $\vec{A_{\text{per}, i}}$ is obtained and an effective distance error $\sigma_{\mathrm{eff},i}$ is calculated. The midpoint $\vec{M_i} =\frac{\vec{E_i} + \vec{J_i}}{2}$ is used to compute the line-of-sight unit vector $\hat{D_i} = \frac{\vec{A_{\text{true}, i}} - \vec{M_i}}{\| \vec{A_{\text{true}, i}} - \vec{M_i} \|}$ from the midpoint position to the true asteroid position. The deviation error vector $\vec{\delta_i} = \vec{A_{\text{per, i}}} - \vec{A_{\text{true, i}}}$ between the perturbed asteroid and the true asteroid position is decomposed into longitudinal $\delta_{\parallel, i} = \vec{\delta_i} \cdot \hat{D_i}$ and transverse $\delta_{\perp,i} = \|\vec{\delta_i} -  (\vec{\delta_{\parallel, i}})   \hat{D}\|$ scalar components. If $\delta_{\parallel, i} > \sigma_{\mathrm{eff},i}$ or $\delta_{\perp,i} > (\delta_{\text{avg}})   \| \vec{A_{\text{true}, i}} - \vec{M_i} \|$ where $\delta_{\text{avg}}$ (defined in equation \ref{eqn:parallaxerror}) is in radians, then the candidate perturbed orbit is rejected for that specific observation set and will continue to be evaluated in the same method for the remaining observation sets. If the error conditions are fulfilled for the specific observation time $i$, the candidate perturbed orbit will be propagated to the next observation time $i+1$ in the observation set where $\vec{\delta_{i+1}}$ will be computed to determine the validity of the orbit once again. If the candidate is within $\delta_{\parallel, i}$ and $\delta_{\perp,i}$ error constraints for all observation times in an observation set, it is accepted as a valid orbit for that observation set and the candidate is then evaluated for the remaining observation sets. This procedure occurs for the 10 million perturbed candidate orbits until all valid orbits are identified for each observation set. The valid orbits are then propagated to the time-of-impact epoch $I$ and the localization error $\sigma_{\|\vec{\delta_I}\|, N}$ for each observation set is computed by evaluating the standard deviation of $\|\vec{\delta_I}\| = \| \vec{A_{\text{per, I}}} - \vec{A_{\text{true, I}}} \|$ of all respective valid orbits.

\section{Results} \label{sec:results}

Table \ref{tab:error-stats} summarizes the results obtained from Figure \ref{fig:hist} where the standard deviation of $\|\vec{\delta_{I}}\|$ (which is equivalently defined as the localization error $\sigma_{\|\vec{\delta_I}\|, N}$) for valid orbits are evaluated for all observation groups. As anticipated, the localization error reduces with each additional observation conducted acquiring a 1-$\sigma$ localization error of $1.42 \ R_\oplus$, smaller than the Earth's diameter, after 8 parallax measurements with the parameters defined in Section \ref{sec:methodology}.\footnote[3]{Conducting the Monte Carlo simulation for 10 observations with 7-day separation achieves a 1-$\sigma$ localization error of $\sigma_{\|\vec{\delta_I}\|}  \approx 1 \ R_\oplus$.}

\begin{table}[h]
    \centering
    \caption{\textit{Standard Deviation $(\sigma)$ and Mean $(\mu)$ of $\|\vec{\delta_I}\|$ Error Values for each Observation Group $(N)$.} Results summarized from Figure \ref{fig:hist} and \ref{fig:prob}.}
    \label{tab:error-stats}
    \resizebox{\columnwidth}{!}{%
    \begin{tabular}{ccc}
        \toprule
        \textbf{Observation Group} & \textbf{Standard Deviation ($\sigma$)} & \textbf{Mean ($\mu$)} \\
        \midrule
        N = 1 & 2.71 $R_\oplus$ & 4.39 $R_\oplus$ \\
        N = 2 & 2.68 $R_\oplus$ & 4.28 $R_\oplus$ \\
        N = 3 & 2.64 $R_\oplus$ & 4.13 $R_\oplus$ \\
        N = 4 & 2.55 $R_\oplus$ & 3.90 $R_\oplus$\\
        N = 5 & 2.38 $R_\oplus$ & 3.54 $R_\oplus$\\
        N = 6 & 2.05 $R_\oplus$ & 3.03 $R_\oplus$\\
        N = 7 & 1.68 $R_\oplus$ & 2.53 $R_\oplus$\\
        N = 8 & 1.42 $R_\oplus$ & 2.17 $R_\oplus$\\
        \bottomrule
    \end{tabular}%
}
\end{table}

\begin{figure}[htb!]
    \centering
    \includegraphics[width=\linewidth]{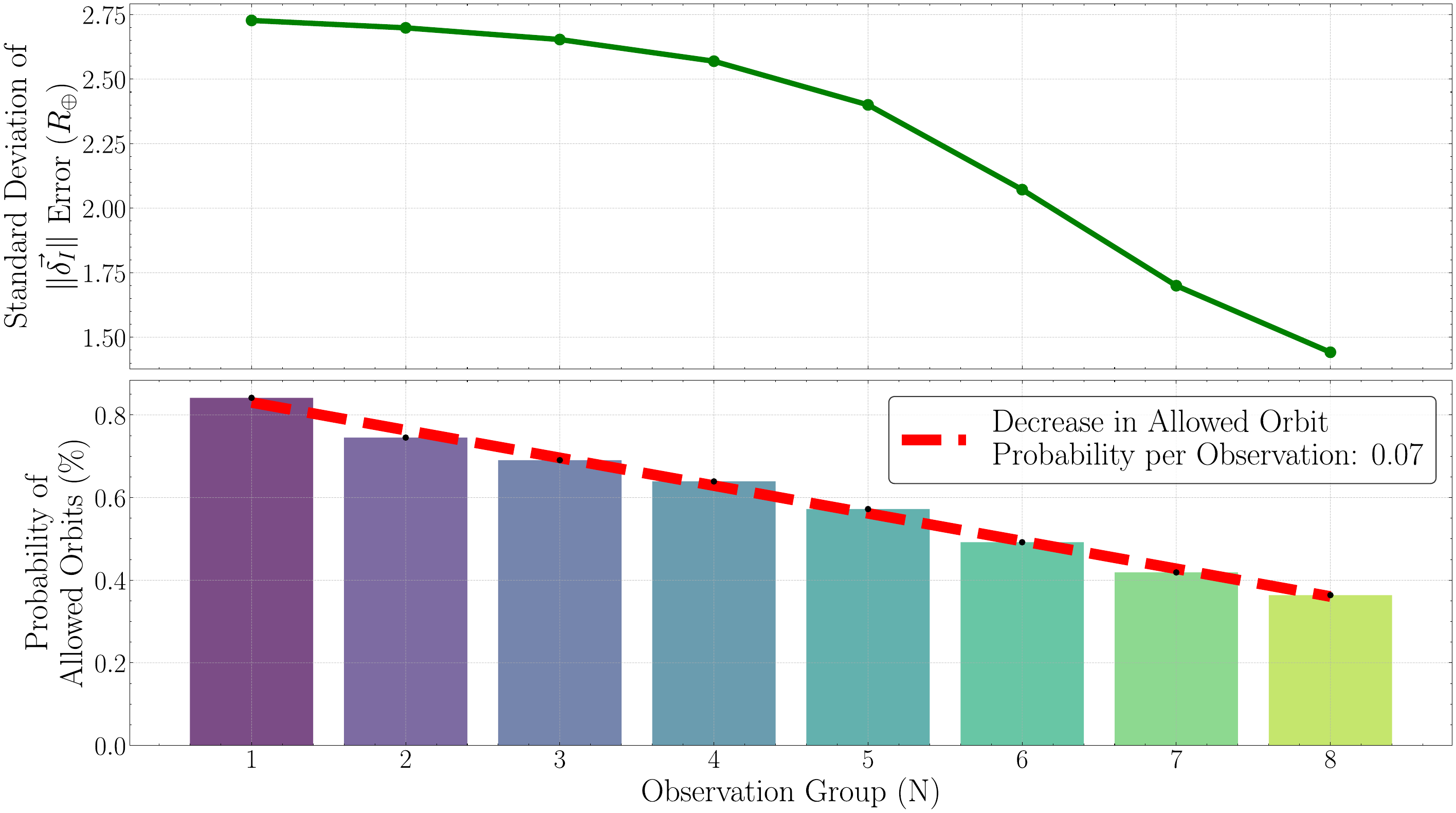}
    \caption{\textit{Localization Error (Standard Deviation of $\|\vec{\delta_I}\|$ Error $[R_\oplus]$) per Observation Group (top). Probability of Allowed Orbits $(\%)$ per Observation Group (bottom).} The probability of allowed orbits is computed by dividing the number of valid orbits for each observation group (see Table \ref{tab:allowed-orbits}) by the total orbits possible (10 million) and taking the percentage. A linear regression line (plotted in dashed red) is fitted to the probability points, determining the decrease in probability percentage per observation group (slope) to be $\approx0.07$.}
    \label{fig:prob}
\end{figure}

Figure \ref{fig:prob} \textit{(top)} details the correlation for the localization error as a function of the number of observations conducted. A linear correlation $(\frac{\partial \sigma_{\|\vec{\delta_I}\|}}{\partial N} \approx 0.3 \ \text{AU})$ exists with steep reductions in localization error within the domain $N \in [4, 8]$. This segment is assumed to be approximately where the inflection point is located of a logistic decline curve which asymptotes to a steady-state critical localization error dependent on the angular uncertainty of the JWST-Earth telescope system $\delta_{\text{avg}}$ and the time-dependent parallax distance error $\sigma_{\mathrm{p}}$ of the NEO-of-concern as it traverses away from JWST-Earth and towards aphelion $(Q)$. 

It is important to note that $\delta_{\text{avg}}$ and consequently $\sigma_{\mathrm{p}}$ can be minimized by utilizing an Earth-based telescope with minimal $\delta_2$ arcsecond uncertainty (see equation \ref{eqn:parallaxerror}) such as the Hubble Space Telescope (HST) with a $\delta_2$ of $\approx 0.007^{\prime\prime}$ \citep{beals1988hubble}. HST and other similar high-precision telescopes should be utilized as the Earth-based telescope in order to obtain more optimal orbital constraints in the scenario of when the NEO-of-concern is particularly hazardous with an elevated likelihood of collision with Earth.

The Y-axis within Figure \ref{fig:prob} (\textit{bottom}) is defined as the probability percentage of a candidate orbit (any within the 10 million generated via Monte Carlo) being accepted as a valid orbit (procedure defined in Section \ref{sec:monte}) for an individual observation set (see Table \ref{tab:allowed-orbits} for explicit results). Figure \ref{fig:prob} (\textit{bottom}) demonstrates the inverse linear relationship between the percentage of allowed orbits $(\%)$ and the number of observations conducted where $0.7\%$ of candidate orbits are eliminated with each additional observation.

\begin{table}[htbp]
    \centering
    \caption{\textit{Amount and Percentage of Allowed Orbits for Each Observation Group.} Results summarized from Figure \ref{fig:prob}.}
    \label{tab:allowed-orbits}
    \resizebox{\columnwidth}{!}{%
    \begin{tabular}{ccc}
        \toprule
        \textbf{Observation Group} & \textbf{Allowed Orbits} & \textbf{Allowed Orbits (\%)} \\
        \midrule
        N = 0 & 10,000,000 & 100.00\% \\
        N = 1 & 84,153     & 0.8415\% \\
        N = 2 & 74,526     & 0.7453\% \\
        N = 3 & 69,057     & 0.6906\% \\
        N = 4 & 63,909     & 0.6391\% \\
        N = 5 & 57,217     & 0.5722\% \\
        N = 6 & 49,190     & 0.4919\% \\
        N = 7 & 41,880     & 0.4188\% \\
        N = 8 & 36,392     & 0.3639\% \\
        \bottomrule
    \end{tabular}%
    }
\end{table}

Table \ref{tab:allowed-orbits} provides the explicit number of allowed orbits for each observation group and the respective probability percentage. With 1 parallax measurement, the number of valid orbits decreases by $\sim 10^6$ with a reduction in probability percentage of $99.1585 \%$. A linear relationship then exists for all subsequent observations where the allowed orbits for a set of 8 observations are constrained to be 36,392 out of $10^7$ potential orbits with a $0.3639\%$ probability.

\section{Conclusions} \label{sec:conc}

We utilize the example of 2024 YR4 to demonstrate the use of JWST in simultaneous observations with an Earth-based telescope for parallax measurements in tightly constraining the orbital trajectory of hazardous NEOs. We identify, via Monte Carlo simulation, a significant reduction in localization error (1-$\sigma$) at the time-of-impact epoch of 2024 YR4 to a standard deviation below the diameter of the Earth with 8 epochs of observations spaced 7 days apart. Employing this method for the forthcoming JWST observation of 2024 YR4 may provide refinement in the probability accuracy of collision with the Moon in 2032 \citep{rivkin2025jwst}.

We emphasize how the proposed parallax method exploits the considerable parallax angles and baselines formed from JWST-Earth-NEO orbit positions and consequently obtains accurate distance measurements, allowing the method to be effective at NEO orbit trajectory constraints. The approach is therefore most optimal for NEO orbital trajectories that create extensive parallax angles at a minimal distance away from the JWST-Earth observer system. Nevertheless, the method still demonstrated effectiveness for localization in the 2024 YR4 instance despite the relatively minimal parallax angles formed while the NEO traversed outward towards aphelion at substantial distances. With the likely detection of numerous NEOs in the coming years by the Vera C. Rubin Observatory, the proposed parallax method could be applied to several new NEOs that pose potential existential risks. 

We note the limitations of this method concerning the operational logistics of JWST where observational use is especially competitive as a consequence of the importance of JWST to many areas of astrophysics research. We hence emphasize the potential necessity for an auxiliary telescope positioned at L2 designed exclusively for NEO detection and localization for planetary defense and national security purposes \citep{loeb2025discovering}. A more comprehensive surveillance and targeting system for hazardous NEOs could exist in the form of NEO-specific telescopes positioned at all Lagrange points thereby eliminating existing NEO blindspots and enabling the continuous observation of hazardous NEOs irrespective of their orbital trajectory in relation to Earth.

\begin{acknowledgments}
We thank the Harvard College Research Program (HCRP), the Galileo Project, and the Black Hole Initiative for support of this work. Sriram tremendously thanks his research advisor, Professor Abraham Loeb, for his never-ending and continuous support, and for providing immense inspiration in the pursuit of exploring untried ideas in science.
\end{acknowledgments}

\software{NASA JPL Horizons System \citep{NASA_JPL_2025},
          SciencePlots \citep{SciencePlots},  
          Matplotlib \citep{Hunter:2007}, 
          Numpy \citep{harris2020array},
          Astropy \citep{astropy:2013, astropy:2018, astropy:2022},
          ChatGPT \citep{OpenAI_2025, achiam2023gpt, hurst2024gpt, OpenAI_2025b},
          DeepSeek \citep{guo2025deepseek}, 
          Anthropic \citep{Anthropic_2025, Anthropic_2025a}
}



\bibliographystyle{aasjournalv7}
\bibliography{sample7}{}




\end{document}